\documentclass{article}
\usepackage{amsmath}
\usepackage{amsthm}
\usepackage{amssymb}
\usepackage{esint}
\usepackage{fullpage}
\usepackage{enumitem}
\allowdisplaybreaks
\makeatletter
\numberwithin{equation}{section}
\numberwithin{figure}{section}
\theoremstyle{plain}
\newtheorem{thm}{Theorem}
\newtheorem*{thm*}{Theorem}
\theoremstyle{definition}
\newtheorem{defn}[thm]{Definition}
\theoremstyle{plain}
\newtheorem{algorithm}{Algorithm}
\theoremstyle{plain}

\theoremstyle{remark}
\newtheorem{rem}[thm]{Remark}
\theoremstyle{plain}

\newtheorem{prob}{Problem}

\newtheorem{notn}[thm]{Notation}
\theoremstyle{definition}
\newtheorem{example}{Example}

\newcommand{\inv}{\mbox{$\mathrm{inv}$}}
\newcommand{\ing}{\mbox{$\mathrm{int}$}}
\newcommand{\Ei}{\mbox{$\mathrm{Ei}$}}
\newcommand{\li}{\mbox{$\mathrm{li}$}}

\begin{document}

\title{Algorithm for computing\\ 
Semi-Fourier sequences of \\
Expressions involving exponentiations and integrations}

\author{Hoon Hong \and  Adam Strzebo\'nski
\thanks{The research of Hoon Hong  was partially supported by the following grants  US NSF 1319632.}}
\renewcommand\footnotemark{}
\maketitle


\begin{abstract}
We provide an algorithm for  computing semi-Fourier
sequences for expressions constructed from arithmetic operations,
exponentiations and integrations. The semi-Fourier sequence is a relaxed
version of Fourier sequence for polynomials (expressions made of additions and
multiplications).
\end{abstract}

\section{Introduction}

The main contribution of this paper is an algorithm for computing semi-Fourier
sequences for expressions constructed from arithmetic operations,
exponentiations and integrations. The semi-Fourier sequence is a relaxed
version of Fourier sequence for polynomials (expressions made of additions and
multiplications). Below we elaborate on the above statements.

Fourier sequence of polynomials is a finite sequence of successive
differentiation. As an example, consider
\[
e=x^{3}+3x^{2}+5x+7
\]
Then the Fourier sequence of $e$ is given by%
\[%
\begin{array}
[t]{lllll}%
g_{1} & = & e & = & x^{3}+3x^{2}+5x+7\\
g_{2} & = & g_{1}^{\prime} & = & 3x^{2}+6x+5\\
g_{3} & = & g_{2}^{\prime} & = & 6x+6\\
g_{4} & = & g_{3}^{\prime} & = & 6
\end{array}
\]
\noindent It has an obvious but important property: the successive
differentiation eventually leads to a constant. It also has another nice
property due to Budan-Fourier~\cite{B,F}: Let $\nu(x)$ denote the number 
of sign changes in 
the sequence $g_{1}(x),\ldots$ (with zeros ignored). Then the number 
of roots of $e$ in $(a, b]$, counted with multiplicities, is equal to
$\nu(a)-\nu(b)-2 s$, where $s$ is a nonnegative integer.

One naturally wonders whether one can do the same for  non-polynomials.
Consider%
\[
e=\exp\left(  x\right)  +\exp\left(  x^{2}\right)
\]
Let us compute the successive differentiation as we did above. We obtain
\[%
\begin{array}
[t]{lllll}%
g_{1} & = & e & = & \exp\left(  x\right)  +\exp\left(  x^{2}\right) \\
g_{2} & = & g_{1}^{\prime} & = & \exp\left(  x\right)  +2x\exp\left(
x^{2}\right) \\
g_{3} & = & g_{2}^{\prime} & = & \exp\left(  x\right)  +2\exp\left(
x^{2}\right)  +4x^{2}\exp\left(  x^{2}\right) \\
& \vdots &  &  &
\end{array}
\]
Note that the successive differentiation will \emph{not} lead to a constant.
Thus Fourier sequence for $\exp\left(  x\right)  -\exp\left(  x^{2}\right)  $
does \emph{not} exist.

In~\cite{S1,S2}, Strzebo\'nski (one of the co-authors of this paper) relaxed the
notion of Fourier sequence to \textquotedblleft semi\textquotedblright-Fourier
sequence. For the above example, a semi-Fourier sequence is given by
\begin{equation*}
\begin{array}[t]{lllllllll}
g_{1} & = & e\cdot h_{1} & = & 1+\exp \left( x^{2}\right) \exp \left(
x\right) ^{-1} &  & h_{1} & = & \exp \left( x\right) ^{-1} \\ 
g_{2} & = & g_{1}^{\prime }\cdot h_{2} & = & 2x-1 &  & h_{2} & = & \exp
\left( x^{2}\right) ^{-1}\exp \left( x\right)  \\ 
g_{3} & = & g_{2}^{\prime }\cdot h_{3} & = & 2 &  & h_{3} & = & 1%
\end{array}%
\end{equation*}%
Note that before/after each differentiation one is allowed to multiply by 
another expression, $h_{i}$, which is non-zero for all values of $x$ in 
the domain of $e$. Of course
when we require that $h_{i}=1$, then it is a Fourier sequence. By allowing
$h_{i}$ to be other than $1$, one obtains a relaxed version of Fourier
sequence. Since $h_{i}$ is non-zero for all real values of $x$,
 the semi-Fourier sequence  still preserves the Budan-Fourier's  
 property mentioned above (Theorem 2.4 in \cite{S1}). 

Strzebo\'nski also proved constructively  that a
semi-Fourier sequence exists for expressions constructed from arithmetic
operations, exponentiations, logarithms and arctangents, by providing  an
algorithm for computing such a sequence.\footnote{The definition of 
a semi-Fourier sequence in this paper is a bit stricter than the definition used in~\cite{S1,S2}, 
however the semi-Fourier sequences constructed in~\cite{S1,S2} satisfy
the stricter requirements.} 

One again naturally wonders whether semi-Fourier sequences exist for even
larger class of functions and whether one can find them algorithmically. One
natural class  to consider is the set (called Exp-Int) of expressions
constructed from arithmetic operations, exponentiations and integrations (\ing). 
It is easy to see that logarithms and arctangents belong to Exp-Int, since they are integrals of $1/x$ and $1/(1+x^2)$.   
However there are many expressions in Exp-Int that can{\em not} be expressed 
in terms of arithmetic operations, exponentiation, logarithms and arctangents, 
such as  
\begin{eqnarray*}
e &=&\exp(x\ \ing(\exp(-x^2))) \;\;-\;\; \ing(\exp(-x^2)) \;\;-\;\; 3\\
  &=&\exp\left(x \frac{\sqrt {\pi }}{2}{\rm erf} \left(x\right)\right)\;\;-\;\;\frac{\sqrt {
\pi }}{2} {\rm erf} \left(x\right)\;\;-\;\;3
\end{eqnarray*}

Many elementary and special functions can be represented as Exp-Int
expressions, such as 
rational functions,
exponential functions,
logarithm,
radicals ($\sqrt[k]{}$),
inverse trigonometric functions ($\arcsin,\arccos,\arctan$),
(inverse) hyperbolic trigonometric functions ($\sinh,\cosh,\tanh,\operatorname{arsinh},\operatorname{arcosh},\operatorname{artanh}$),
hyperbolic sine/cosine integrals $(\operatorname{Shi}, \operatorname{Chi})$,
logistic sigmoid function ($F$),
error function ($\operatorname{erf}$),
logarithmic integral function ($\operatorname{li}$),
exponential integral function ($\operatorname{Ei}$),
polylogarithm function ($\operatorname{Li}_s$),
Spence function ($\operatorname{Li}_{2}$),
Gudermannian function ($\operatorname{gd}$),
Dawson function ($D_{\pm}$),
Kummer function ($\Lambda_{n}$),
Incomplete beta function ($B_{ab}$),
Incomplete gamma function ($\Gamma_{s}$),
Incomplete elliptic integral ($F_{k}$), etc.

The main contribution of this paper is to prove constructively that
there exists a semi-Fourier sequence for every Exp-Int expression, by providing  an
algorithm for computing such a sequence.  For a semi-Fourier sequence computed 
by the algorithm  for the above $e$, see Example~\ref{ex1}.  

Of course, the main task of such an algorithm is
figuring out $h_i$'s  at each step, so that the sequence eventually terminates.
It was quite challenging because most choices of $h_i$'s lead to infinite sequences
and because, at each step, it was by no means obvious to predict which $h_i$'s would be the lucky ones. We overcame the challenge 
by defining a partial ordering (ranking) on expressions, and ensuring that the rank decreases.

One again wonders whether there is even larger class of functions for which 
semi-Fourier sequences exist. It is easy to see that semi-Fourier sequences 
exist for any real analytic function $e$ with finitely many roots, since one 
can take $g_1$ to be a polynomial which has the same roots as $e$ with 
the same multiplicities. However, in general $g_1$ cannot be constructed
algorithmically, since we cannot compute the roots of $e$ (in fact, 
the original reason for introducing semi-Fourier sequences was to be
able to isolate roots algorithmically). Finding a larger class of functions
for which semi-Fourier sequences exist and can be found algorithmically
is an open problem which we leave for future work.

There are several closely related works.
As mentioned above, in~\cite{S1,S2}, Strzebo\'nski, one of the authors of the present paper,
introduced the notions, and algorithms for semi-Fourier sequence for
exponential - logarithmic - arctan expressions and showed that  the sequences  have the Budan-Fourier property.   This present paper  gives an algorithm for larger class of expressions. By construction, it  is immediate that the output sequences have the Budan-Fourier property.   In \cite{S1,S2}, Strzebo\'nski also gave real root isolation algorithms. However, in the present paper, we do not do so, mainly because  such real root isolation algorithm would require zero-testing which is non-trivial for Exp-Int  and interesting on its own, deserving a separate paper.  Hence we leave it for the future research. 
    
In~\cite{R}, Richardson gave an algorithm for  recognizing zeros among exponential-logarithmic expressions. For this, he generates certain sequences of expressions.  They are similar but not the same as a semi-Fourier sequence. Thus  they do not have the Budan-Fourier property either. This present paper  deals with
larger family of expressions and the results have  the Budan-Fourier property.

In~\cite{K}, Kovanskii studied zeros of 
Pfaffian functions.  It involves sequence of  functions defined by certain differential equations.    The sequence is not semi-Fourier in general. As the result, it does not have the Budan-Fourier property.  The present paper deals with smaller family of expressions. However, it generates semi-Fourier sequence with the Budan-Fourier property.

The paper is structured as follows:
In Section~\ref{problem}, we give a precise statement of the problem that will be tackled in this paper.  
In Section~\ref{algorithm}, we provide an algorithm solving the problem.
In Section~\ref{correct}, we prove the correctness of the algorithm.
In Section~\ref{examples}, we provide several examples.

\section{Problem}
\label{problem}

In this section, we state the main problem precisely. For this, we need
several notions.

\begin{defn}[Exp-Int expression]
An \emph{exp-int expression } is an expression constructed with the
variable symbol $x$, rational numbers, the binary operator symbols $+$ and $%
\cdot$ and the unary operator symbols $\mbox{$\mathrm{inv}$}$, $\exp$, and $%
\mbox{$\mathrm{int}$}$. The set of all exp-int expressions is denoted by $EI$.
\end{defn}
Of course, the symbols $\mathrm{inv}$, $\mathrm{exp}$ and $\mathrm{int}$ stand for inverse, exponential and  integration (anti-derivative) respectively.  
\noindent As usual, for convenience, we use the following usual short-hands: 
$$f^{n}=\underset{n}{\underbrace{f\cdots f}}\;\;\;\;\;\; 
  f^{-n}=\mathrm{inv}(f)^{n}\;\;\;\;\;\;
  f^{0}=1\;\;\;\;\;\;
  -e=-1\cdot e$$

\begin{example}
We list several expressions that can be  rewritten as   Exp-Int expressions.
\begin{itemize}
\item $e^x \left(e^{\frac{1}{x}-e^{-x}}-e^{\frac{1}{x}}\right)+5$

      $\exp (x) (\exp (\inv(x)-\exp (-x))-\exp (\inv(x)))+5$ 

\item $e^x \log x + e^{x^2}+x$

      $\exp(x)\  \ing(\inv(x)) +\exp(x^2)+x$
      
\item $ \frac{x}{\exp (x)-1}-\log (1-\exp (-x))-\frac{1}{x}$

      $x \inv(\exp (x)-1)-\ing(\inv(\exp (x)-1))-\inv(x)$

\item $ 2 \arccos\left(\frac{x}{2}\right)-\frac{\pi }{2}-\frac{1}{2} x \sqrt{4-x^2}$

      $2\, \ing\left(-\exp \left(-\frac{1}{2}\, \ing\left(-2\, x\, \inv\left(4-x^2\right)\right)\right)\right)-\frac{1}{2} \, x \exp \left(\frac{1}{2}\, \ing\left(-2 \, x \, \inv\left(4-x^2\right)\right)\right) $

\item $\Ei\left(x^2+2\right)-\, \li\left(x^2+2\right)-e^{x^2+2}$

      $\ing\left(2 x \exp \left(x^2+2\right) \, \inv\left(x^2+2\right)\right)-\, \ing\left(2 x \, \inv\left(\, \
\ing\left(2 x \, \inv\left(x^2+2\right)\right)\right)\right) -\exp \left(x^2+2\right)$
\end{itemize}
In the above, we have used the following translations. Note that integrals
may include arbitrary constants.
\begin{eqnarray*}
\log(x) &=&\int \frac{1}{x} \\
\sqrt[n]{x} &=&e^{\frac{1}{n}\log (x)} \\
\arccos(x) &=&\int -\frac{1}{\sqrt{1-x^2}} \\
\Ei(x) &=&\int \frac{e^{x}}{x} \\
\li(x) &=&\int \frac{1}{\log \left( x\right) }
\end{eqnarray*}\end{example}

\begin{defn}[Expression differentiation]
\emph{Expression differentiation} is the operation $D:EI\rightarrow EI$
defined recursively as follows.

\begin{enumerate}
\item $D(x)=1$,

\item $D(c)=0$, for $c\in\mathbb{Q}$,

\item $D(f+g)=D(f)+D(g)$,

\item $D(f\cdot g)=D(f)\cdot g+f\cdot D(g)$,

\item $D(f^{-1})=- f^{-2}\cdot D(f)$,

\item $D(\exp(f))=\exp(f)\cdot D(f)$,

\item $D(\mbox{$\mathrm{int}$}(f))=f$
\end{enumerate}
\end{defn}

\begin{rem}
We explicitly listed the above well known properties of differentiation,
because we want to emphasize the obvious but crucial facts (that will be
used later):

\begin{itemize}
\item The derivative (the result of the differentiation) of an exp-int
expression is itself an exp-int expression.

\item Let $T \in \{\mbox{$\mathrm{inv}$},\mbox{$\mathrm{int}$},\exp\}$. If $%
T(g)$ is a subexpression of $D(f)$, then it is a subexpression of $f$.
\end{itemize}
\end{rem}

\begin{rem}
An expression can be naturally viewed as a partial function, after fixing
the meaning of the distinct int subexpressions by choosing anti-derivatives.
\end{rem}

\break

\begin{notn}
We write $f \rightarrowtail g$ iff for every choice of anti-derivatives
corresponding to the distinct int subexpressions in $f$ and $g$ and for
every $x$ in the domain of $f$, we have $f(x)=g(x)$.
\end{notn}

\begin{defn}[Semi-Fourier sequence]
\label{DefSF} Let $e\in EI$. A sequence $(g_{1},h_{1}),\ldots,(g_{m},h_{m})$
of pairs of elements of $EI$ is a \emph{semi-Fourier sequence} for $e$ if

\begin{enumerate}
\item $e\rightarrowtail g_{1}h_1^{-1}$

\item $D(g_{k})\rightarrowtail g_{k+1}h_{k+1}^{-1}$, for $1\leq k<m$,

\item $D(g_{m})\rightarrowtail0$.
\end{enumerate}
\end{defn}

\noindent Now we are ready to state the problem precisely.
\begin{prob}[Main] Devise an algorithm with the following specification.
\begin{description}[leftmargin=4em,style=nextline,itemsep=0.3em]
\item[\sf Input:]  $e \in EI$
\item[\sf Output:] a semi-Fourier sequence of $e$
\end{description}
\end{prob}

\begin{example}\label{ex1} \
\begin{description}[leftmargin=4em,style=nextline,itemsep=0.3em]
\item[\sf Input:]  $e=\exp(x\ \ing(\exp(-x^2))) - \ing(\exp(-x^2)) - 3$
\item[\sf Output:]
$
\begin{array}[t]{r|l|l} 
i & g_i & h_i \\ 
\hline 
1 & f_3-f_2-3 & 1 \\ 
2 & \left(f_1^{-1} f_2+x\right) f_3-1 & f_1^{-1} \\ 
3 & f_2^2+\left(2 \left(x f_1\right)+2 x\right) f_2+x^2 f_1^2+2 f_1 & f_3^{-1} f_1 \\ 
4 & \left(\left(-4 x^2+4\right) f_1+2\right) f_2+\left(-4 x^3+4 x\right) f_1^2-2 \left(x f_1\right) & 1 \\ 
5 & \left(8 x^3-16 x\right) f_2+\left(16 x^4-32 x^2+8\right) f_1+4 x^2 & f_1^{-1} \\ 
6 & \left(24 x^2-16\right) f_2+\left(-32 x^5+136 x^3-96 x\right) f_1+8 x & 1 \\ 
7 & 48 \left(x f_2\right)+\left(64 x^6-432 x^4+624 x^2-112\right) f_1+8 & 1 \\ 
8 & 48 f_2+\left(-128 x^7+1248 x^5-2976 x^3+1520 x\right) f_1 & 1 \\ 
9 & 256 x^8-3392 x^6+12192 x^4-11968 x^2+1568 & f_1^{-1} \\ 
10 & 2048 x^7-20352 x^5+48768 x^3-23936 x & 1 \\ 
11 & 14336 x^6-101760 x^4+146304 x^2-23936 & 1 \\ 
12 & 86016 x^5-407040 x^3+292608 x & 1 \\ 
13 & 430080 x^4-1221120 x^2+292608 & 1 \\ 
14 & 1720320 x^3-2442240 x & 1 \\ 
15 & 5160960 x^2-2442240 & 1 \\ 
16 & 10321920 x & 1 \\ 
17 & 10321920 & 1 \\ 
\end{array}
$

\noindent where
$
\begin{array}[t]{lll}
f_1 &=& \exp(-x^2)\\
f_2 &=& \ing(f_1)\\
f_3 &=& \exp(x f_2)
\end{array}
$
\end{description}
\end{example}

\break

\section{Algorithm}
\label{algorithm}
In this section, we describe an  algorithm that solves the problem posed in 
the previous section.     
 
\begin{defn}[Equivalence]
Let $f,g\in EI$. We say that $f$ and $g$ are \emph{equivalent}, and write $%
f\sim g$, if $f$ can be transformed into $g$ by applying the commutative
ring properties of $+$ and $\cdot$. We write $f\lesssim g$ if $f$ is
equivalent to a subexpression of $g$.
\end{defn}

\begin{example}
\hfill

\begin{itemize}
\item $(x+1)^2\sim x^2+2 x+1$

\item $\mbox{$\mathrm{inv}$}(x)\lesssim x+\mbox{$\mathrm{int}$}(%
\mbox{$\mathrm{inv}$}(x))$
\end{itemize}
\end{example} 
\begin{defn}[Extension tower]
\label{DefExtTower} An \emph{extension tower}
$E\langle f_{1},\ldots,f_{r}\rangle\subseteq EI$, where $r\geq0$,
is defined recursively as follows
\begin{enumerate}
\item $E\langle\rangle=\mathbb{Q}[x]$,
\item For $1\leq k\leq r$, $f_{k}=T(p_{k})$, 
$p_{k}\in E\langle f_{1},\ldots,f_{k-1}\rangle$, and 
\[
E\langle f_{1},\ldots,f_{k}\rangle
= \left\{
\begin{array}{lll}
\{a_{n}f_{k}^{n}+\ldots+a_{0}\::\: a_{0},\ldots,a_{n}\in 
E\langle f_{1},\ldots,f_{k-1}\rangle\}
& if & T=\inv \\
\{a_{n}f_{k}^{n}+\ldots+a_{0}\::\: a_{0},\ldots,a_{n}\in 
E\langle f_{1},\ldots,f_{k-1}\rangle\}
& if & T=\ing \\
\{f_{k}^{u}(a_{n}f_{k}^{n}+\ldots+a_{0})\::\: 
a_{0},\ldots,a_{n}\in E\langle f_{1},\ldots,f_{k-1}\rangle\}
& if & T=\exp
\end{array}
\right.
\] 
where $a_{n}\neq 0$ and if $T=\exp$ then $a_{0}\neq 0$ and $u\in\mathbb{Z}$.
\end{enumerate}
\end{defn}

\noindent Note that if $f,g\in E\langle f_{1},\ldots,f_{r}\rangle$ then $f+g$, $fg$, 
and $D(f)$ can be easily transformed into equivalent elements of
$E\langle f_{1},\ldots,f_{r}\rangle$. In the following when we write 
$f+g$, $fg$, and $D(f)$ for expressions 
$f,g\in E\langle f_{1},\ldots,f_{r}\rangle$ we mean equivalent 
elements of $E\langle f_{1},\ldots,f_{r}\rangle$. 

\begin{defn}[Rank]
The \emph{rank} of an exp-integrate expression is defined recursively
as follows:
\begin{enumerate}
\item $rank(x)=0$,
\item $rank(c)=0$ for $c\in\mathbb{Q}$,  
\item $rank(f+g)=\max(rank(f),rank(g))$,
\item $rank(f\cdot g)=\max(rank(f),rank(g))$,
\item $rank(\inv(f))=rank(f)+1$,
\item $rank(\exp(f))=rank(f)+1$,
\item $rank(\ing(f))=rank(f)+1$.
\end{enumerate}
\end{defn}

\break

\begin{algorithm}[$EISF$] \label{AlgEISF}{--- \rm\bf Main} ---
\begin{description}[leftmargin=4em,style=nextline,itemsep=0.3em]
\item[\sf Input:]  $e \in EI$
\item[\sf Output:] a semi-Fourier sequence of $e$
\end{description}
\begin{enumerate}[leftmargin=!,labelwidth=1em]
\item Find a sequence $e_{1},\ldots,e_{r}$ of all distinct subexpressions
of $e$ of the form $T_{i}(u_{i})$, where $T_{i}\in\{\inv,\ing,\exp\}$,
ordered by the non-decreasing value of $rank$.
\item Set $f;f_{1},\ldots,f_{r}\leftarrow ET(e;e_{1},\ldots,e_{r})$.
\item Return $ETSF(f;f_{1},\ldots,f_{r})$.
\end{enumerate}
\end{algorithm}

\begin{algorithm}[$ET$] \label{AlgExtTower}\
\begin{description}[leftmargin=4em,style=nextline,itemsep=0.3em]
\item[\sf Input:]   
$a_{1},\ldots,a_{m};e_{1},\ldots,e_{r}$ such that 
\begin{itemize}
\item $a_{1},\ldots,a_{m}\in EI$
\item $e_{1},\ldots,e_{r}$ is a
sequence containing all distinct subexpressions of $a_{1},\ldots,a_{m}$
of the form $T_{i}(u_{i})$, where $T_{i}\in\{\inv,\exp,\ing\}$, ordered
by the increasing value of $rank$.
\end{itemize} 
\item[\sf Output:]
$b_{1},\ldots,b_{m};f_{1},\ldots,f_{r}$ such that
\begin{itemize} 
\item $f_{i}\sim e_{i}$ for $1\leq i\leq r$ 
\item $b_{j}\sim a_{j}$ for $1\leq j\leq m$ 
\item $b_{1},\ldots,b_{m}\in E\langle f_{1},\ldots,f_{r}\rangle$.
\end{itemize}
\end{description}

\begin{enumerate}[leftmargin=!,labelwidth=1em]

\item If $r=0$

\begin{enumerate}
\item For $1\leq j\leq m$ rewrite $a_j$ as   $a_j \sim  c_{j,n_{j}}x^{n_{j}}+\ldots+c_{j,0}=b_{j}$.
\item For $1\leq j\leq m$ set $b_j \leftarrow  c_{j,n_{j}}x^{n_{j}}+\ldots+c_{j,0}$.
\item Return $(b_{1},\ldots,b_{m};)$.
\end{enumerate}
\item If $e_{r}=\inv(v_{r})$ 

\begin{enumerate}
\item For $1\leq j\leq m$ rewrite $a_{j}$ as
$a_{j}\sim c_{j,n_{j}}e_{r}^{n_{j}}+\ldots+c_{j,0}$, where $c_{j,k}$
are free of $e_{r}$, for $0\leq k\leq n_{j}$.

\item Compute $
(d_{1,0},\ldots,d_{m,n_{m}},w_{r};f_{1},\ldots,f_{r-1})=ET(c_{1,0},\ldots,c_{m,n_{m}},v_{r};e_{1},\ldots,e_{r-1})$.

\item Set $f_{r}\leftarrow \inv(w_{r})$.\item For $1\leq j\leq m$, set $b_{j}\leftarrow d_{j,n_{j}}f_{r}^{n_{j}}+\ldots+d_{j,0}$.
\item Return $(b_{1},\ldots,b_{m};f_{1},\ldots,f_{r})$.
\end{enumerate}

\item If $e_{r}=\ing(v_{r})$ 

\begin{enumerate}
\item For $1\leq j\leq m$ rewrite $a_{j}$ as
$a_{j}\sim c_{j,n_{j}}e_{r}^{n_{j}}+\ldots+c_{j,0}$, where $c_{j,k}$
are free of $e_{r}$, for $0\leq k\leq n_{j}$.

\item Compute $
(d_{1,0},\ldots,d_{m,n_{m}},w_{r};f_{1},\ldots,f_{r-1})=ET(c_{1,0},\ldots,c_{m,n_{m}},v_{r};e_{1},\ldots,e_{r-1})$.

\item Set $f_{r}\leftarrow \ing(w_{r})$.\item For $1\leq j\leq m$, set $b_{j}\leftarrow d_{j,n_{j}}f_{r}^{n_{j}}+\ldots+d_{j,0}$.
\item Return $(b_{1},\ldots,b_{m};f_{1},\ldots,f_{r})$.
\end{enumerate}

\item If $e_{r}=\exp(v_{r})$ 

\begin{enumerate}
\item For $1\leq j\leq m$ rewrite $a_{j}$ as $a_{j}\sim e_{r}^{u_{r}}(c_{j,n_{j}}e_{r}^{n_{j}}+\ldots+c_{j,0})$,
where $c_{j,0}\neq0$ and $c_{j,k}$ are free of $e_{r}$, for $0\leq k\leq n_{j}$.

\item Compute $
(d_{1,0},\ldots,d_{m,n_{m}},w_{r};f_{1},\ldots,f_{r-1})=ET(c_{1,0},\ldots,c_{m,n_{m}},v_{r};e_{1},\ldots,e_{r-1})$.

\item Set $f_{r}\leftarrow \exp(w_{r})$.\item For $1\leq j\leq m$, set $b_{j}\leftarrow d_{r}^{u_{r}}(d_{j,n_{j}}f_{r}^{n_{j}}+\ldots+d_{j,0})$.
\item Return $(b_{1},\ldots,b_{m};f_{1},\ldots,f_{r})$.
\end{enumerate}
\end{enumerate}
\end{algorithm}

\break

\begin{algorithm}[$ETSF$] \label{AlgETSF} \ 
\begin{description}[leftmargin=4em,style=nextline,itemsep=0.3em]
\item[\sf Input:]  $f;f_{1},\ldots,f_{r}$ such that $f\in E\langle f_{1},\ldots,f_{r}\rangle$.
\item[\sf Output:] a semi-Fourier sequence $(g_{1},h_{1}),\ldots,(g_{m},h_{m})$
of $f$ such that $g_{i},h_{i}\in E\langle f_{1},\ldots,f_{r}\rangle$,
for $1\leq i\leq m$.
\end{description}
\begin{enumerate}[leftmargin=!,labelwidth=1em]
\item If $f$ does not contain any of $f_{1},\ldots,f_{r}$,

\begin{enumerate}
\item Set $g_{1} \leftarrow f=a_{n}x^{n}+\cdots+a_{0}\in\mathbb{Q}[x]$ and $h_{1}=1$.
\item For $1\leq i\leq n$ set $g_{i+1}\leftarrow D(g_{i})$ and $h_{i+1}\leftarrow1$.
\item \label{o1} Return $(g_{1},h_{1}),\ldots,(g_{n+1},h_{n+1})$.
\end{enumerate}
\item Let $k$ be maximal such that $f_{k}\precsim f$.

\item If $f_{k}=\inv(g)$, let $f=a_{n}f_{k}^{n}+\ldots+a_{0}$.

\begin{enumerate}
\item Set $t \leftarrow a_{0}g^{n}+\cdots+a_{n}$.
\item\label{r1}  Compute $(g_{1},h_{1}),\ldots,(g_{m},h_{m}) \leftarrow ETSF(t;f_{1},\ldots,f_{k-1})$.
\item \label{o2} Return $(g_{1},g^{n}h_{1}),\ldots,(g_{m},h_{m})$.
\end{enumerate}

\item If $f_{k}=\ing(g)$, let $f=a_{n}f_{k}^{n}+\ldots+a_{0}$.

\begin{enumerate}
\item\label{r4} Compute $(e_{1},h_{1}),\ldots,(e_{m},h_{m}) \leftarrow ETSF(a_{n};f_{1},\ldots,f_{k-1})$.
\item For $0\leq i<n$, set $c_{i,1} \leftarrow a_{i}$.
\item For $1\leq j\leq m$,

\begin{enumerate}
\item Set $g_{j}\leftarrow e_{j}f_{k}^{n}+c_{n-1,j}h_{j}f_{k}^{n-1}+\cdots+c_{0,j}h_{j}$.
\item For $0\leq i<n-1$, set $c_{i,j+1} \leftarrow D(c_{i,j}h_{j})+(i+1)c_{i+1,j}h_{j}g$.
\item Set $c_{n-1,j+1} \leftarrow D(c_{n-1,j}h_{j})+ne_{j}g$.
\end{enumerate}
\item Set $i\leftarrow n-1$. While $i\geq0$ and $c_{i,m+1}=0$ decrement $i$.
\item \label{o5} If $i=-1$, return $(g_{1},h_{1}),\ldots,(g_{m},h_{m})$.
\item Set $t \leftarrow c_{i,m+1}f_{k}^{i}+\cdots+c_{0,m+1}$.
\item\label{r5} Compute $(p_{1},q_{1}),\ldots,(p_{l},q_{l}) \leftarrow ETSF(t;f_{1},\ldots,f_{k})$.
\item \label{o6} Return $(g_{1},h_{1}),\ldots,(g_{m},h_{m}),(p_{1},q_{1}),\ldots,(p_{l},q_{l})$.
\end{enumerate}

\item If $f_{k}=\exp(g)$, let $f=f_{k}^{u}(a_{n}f_{k}^{n}+\ldots+a_{0})$.

\begin{enumerate}
\item\label{r2} Compute $(e_{1},h_{1}),\ldots,(e_{m},h_{m}) \leftarrow ETSF(a_{0};f_{1},\ldots,f_{k-1})$.
\item For $1\leq i\leq n$, set $c_{i,1} \leftarrow a_{i}$.
\item For $1\leq j\leq m$,

\begin{enumerate}
\item Set $g_{j} \leftarrow c_{n,j}h_{j}f_{k}^{n}+\cdots+c_{1,j}h_{j}f_{k}+e_{j}$.
\item For $1\leq i\leq n$, set $c_{i,j+1} \leftarrow D(c_{i,j}h_{j})+ic_{i,j}h_{j}D(g)$.
\end{enumerate}
\item Set $i\leftarrow1$. While $i\leq n$ and $c_{i,m+1}=0$ increment $i$.
\item \label{o3} If $i=n+1$, return $(g_{1},f_{k}^{-u}h_{1}),\ldots,(g_{m},h_{m})$.
\item Set $t \leftarrow  c_{n,m+1}f_{k}^{n-i}+\cdots+c_{i,m+1}$.
\item\label{r3} Compute $(p_{1},q_{1}),\ldots,(p_{l},q_{l}) \leftarrow ETSF(t;f_{1},\ldots,f_{k})$.
\item \label{o4} Return $(g_{1},f_{k}^{-u}h_{1}),\ldots,(g_{m},h_{m}),(p_{1},f_{k}^{-i}q_{1}),\ldots,(p_{l},q_{l})$.
\end{enumerate}

\end{enumerate}
\end{algorithm}

\begin{example}\label{ex2} We  illustrate the algorithm $EISF$  by tracing it on the input $e$ from Example~\ref{ex1}.
\[e=\exp(x\ \ing(\exp(-x^2))) - \ing(\exp(-x^2)) - 3\]

\begin{enumerate}
\item 
$
\begin{array}[t]{lll}
e_1 &=& \exp(-x^2)\\
e_2 &=& \ing(\exp(-x^2))\\
e_3 &=& \exp(x \ing(\exp(-x^2)))
\end{array}
$
\item Calling $ET$ with $(e;e_1,e_2,e_3)$, we obtain 

$
\begin{array}[t]{lll}
f   &=& f_3-f_2-3 \\
f_1 &=& \exp(-x^2)\\
f_2 &=& \ing(f_1)\\
f_3 &=& \exp(x f_2)
\end{array}
$
\item Calling $ETSF$ with $(f;f_1,f_2,f_3)$, we obtain

$
\begin{array}[t]{r|l|l} 
i & g_i & h_i \\ 
\hline 
1 & f_3-f_2-3 & 1 \\ 
2 & \left(f_1^{-1} f_2+x\right) f_3-1 & f_1^{-1} \\ 
3 & f_2^2+\left(2 \left(x f_1\right)+2x\right) f_2+x^2 f_1^2+2 f_1 & f_3^{-1} f_1 \\ 
4 & \left(\left(-4 x^2+4\right) f_1+2\right) f_2+\left(-4 x^3+4 x\right) f_1^2-2 \left(x f_1\right) & 1 \\ 
5 & \left(8 x^3-16 x\right) f_2+\left(16 x^4-32 x^2+8\right) f_1+4 x^2 & f_1^{-1} \\ 
6 & \left(24 x^2-16\right) f_2+\left(-32 x^5+136 x^3-96 x\right) f_1+8 x & 1 \\ 
7 & 48 \left(x f_2\right)+\left(64 x^6-432 x^4+624 x^2-112\right) f_1+8 & 1 \\ 
8 & 48 f_2+\left(-128 x^7+1248 x^5-2976 x^3+1520 x\right) f_1 & 1 \\ 
9 & 256 x^8-3392 x^6+12192 x^4-11968 x^2+1568 & f_1^{-1} \\ 
10 & 2048 x^7-20352 x^5+48768 x^3-23936 x & 1 \\ 
11 & 14336 x^6-101760 x^4+146304 x^2-23936 & 1 \\ 
12 & 86016 x^5-407040 x^3+292608 x & 1 \\ 
13 & 430080 x^4-1221120 x^2+292608 & 1 \\ 
14 & 1720320 x^3-2442240 x & 1 \\ 
15 & 5160960 x^2-2442240 & 1 \\ 
16 & 10321920 x & 1 \\ 
17 & 10321920 & 1 \\ 
\end{array}
$
\end{enumerate}
\end{example}

\begin{example}
We illustrate the algorithm $ETSF$  by tracing it on the input $(f;f_1,f_2,f_3)$, which is taken from  Step 3 of Example~\ref{ex2}. 
\[
\begin{array}[t]{lll}
f   &=& f_3-f_2-3 \\
f_1 &=& \exp(-x^2)\\
f_2 &=& \ing(f_1)\\
f_3 &=& \exp(x f_2)
\end{array}
\]
The algorithm $ETSF$ is recursive. Hence we trace the recursive calls.
\begin{small}

\[
\begin{array}{l} 
\texttt{In}_1\texttt{ } = \begin{array}[t]{l}
f_3-f_2-3\end{array} \\ 

|\;\;\;\;\texttt{In}_2\texttt{ } = \begin{array}[t]{l}
-f_2-3\end{array} \\ 

|\;\;\;\;|\;\;\;\;\texttt{In}_3\texttt{ } = \begin{array}[t]{l}
-f_1\end{array} \\ 

|\;\;\;\;|\;\;\;\;\texttt{Out}_3 = \begin{array}[t]{l|l}
-1 & f_1^{-1}
\end{array} \\ 

|\;\;\;\;\texttt{Out}_2 = \begin{array}[t]{l|l}
-f_2-3 & 1 \\ 
-1 & f_1^{-1}
\end{array} \\ 

|\;\;\;\;\texttt{In}_4\texttt{ } = \begin{array}[t]{l}
f_1^{-1} f_2^2+\left(\left(2+2 f_1^{-1}\right) x\right) f_2+x^2 f_1+2\end{array} \\ 

|\;\;\;\;|\;\;\;\;\texttt{In}_5\texttt{ } = \begin{array}[t]{l}
f_1^{-1}\end{array} \\ 

|\;\;\;\;|\;\;\;\;\texttt{Out}_5 = \begin{array}[t]{l|l}
1 & f_1
\end{array} \\ 

|\;\;\;\;|\;\;\;\;\texttt{In}_6\texttt{ } = \begin{array}[t]{l}
\left(\left(-4 x^2+4\right) f_1+2\right) f_2+\left(-4 x^3+4 x\right) f_1^2-2 \left(x f_1\right)\end{array} \\ 

|\;\;\;\;|\;\;\;\;|\;\;\;\;\texttt{In}_7\texttt{ } = \begin{array}[t]{l}
\left(-4 x^2+4\right) f_1+2\end{array} \\ 

|\;\;\;\;|\;\;\;\;|\;\;\;\;\texttt{Out}_7 = \begin{array}[t]{l|l}
\left(-4 x^2+4\right) f_1+2 & 1 \\ 
8 x^3-16 x & f_1^{-1} \\ 
24 x^2-16 & 1 \\ 
48 x & 1 \\ 
48 & 1
\end{array} \\ 

|\;\;\;\;|\;\;\;\;|\;\;\;\;\texttt{In}_8\texttt{ } = \begin{array}[t]{l}
\left(256 x^8-3392 x^6+12192 x^4-11968 x^2+1568\right) f_1\end{array} \\ 

|\;\;\;\;|\;\;\;\;|\;\;\;\;\texttt{Out}_8 = \begin{array}[t]{l|l}
256 x^8-3392 x^6+12192 x^4-11968 x^2+1568 & f_1^{-1} \\ 
2048 x^7-20352 x^5+48768 x^3-23936 x & 1 \\ 
14336 x^6-101760 x^4+146304 x^2-23936 & 1 \\ 
86016 x^5-407040 x^3+292608 x & 1 \\ 
430080 x^4-1221120 x^2+292608 & 1 \\ 
1720320 x^3-2442240 x & 1 \\ 
5160960 x^2-2442240 & 1 \\ 
10321920 x & 1 \\ 
10321920 & 1
\end{array} \\ 

|\;\;\;\;|\;\;\;\;\texttt{Out}_6 = \begin{array}[t]{l|l}
\left(\left(-4 x^2+4\right) f_1+2\right) f_2+\left(-4 x^3+4 x\right) f_1^2-2 \left(x f_1\right) & 1 \\ 
\left(8 x^3-16 x\right) f_2+\left(16 x^4-32 x^2+8\right) f_1+4 x^2 & f_1^{-1} \\ 
\left(24 x^2-16\right) f_2+\left(-32 x^5+136 x^3-96 x\right) f_1+8 x & 1 \\ 
48 \left(x f_2\right)+\left(64 x^6-432 x^4+624 x^2-112\right) f_1+8 & 1 \\ 
48 f_2+\left(-128 x^7+1248 x^5-2976 x^3+1520 x\right) f_1 & 1 \\ 
\multicolumn{2}{l}{\texttt{Out}_8}  \\
\end{array} \\

|\;\;\;\;\texttt{Out}_4 = \begin{array}[t]{l|l}
f_2^2+\left(2 \left(x f_1\right)+2 x\right) f_2+x^2 f_1^2+2 f_1 & f_1 \\ 
\left(\left(-4 x^2+4\right) f_1+2\right) f_2+\left(-4 x^3+4 x\right) f_1^2-2 \left(x f_1\right) & 1 \\ 
\multicolumn{2}{l}{\texttt{Out}_6} 
\end{array} \\ 

\texttt{Out}_1 = \begin{array}[t]{l|l}
f_3-f_2-3 & 1 \\ 
\left(f_1^{-1} f_2+x\right) f_3-1 & f_1^{-1} \\ 
f_2^2+\left(2 \left(x f_1\right)+2 x\right) f_2+x^2 f_1^2+2 f_1 & f_3^{-1} f_1 \\ 
\multicolumn{2}{l}{\texttt{Out}_6} 
\end{array} \\ 

\end{array}
\]

\end{small}
\end{example}

\section{Proof of Correctness of Algorithm}
\label{correct}
In this section, we prove that the main algorithm $EISF$ is correct. 
We need to show the termination and the partial correctness.
 In the following subsections, we will show them one by one.
\subsection{Termination}
Since the main algorithm consists of a fixed sequence of steps,
it suffices to show that the subalgorithms~$ET$ and $ETSF$ terminate. 
The subalgorithm~$ET$ terminates since each recursion reduces the value of 
the natural number $r$ until it becomes $0$. 
Hence, it remains to show that the subalgorithm~$ETSF$ terminates.
For this, it is convenient to introduce an ordering over 
$E\langle f_{1},\ldots ,f_{r}\rangle$.  
\begin{defn}[Ordering]
Let $d$ be the function from $E\langle f_{1},\ldots ,f_{r}\rangle$ to 
$\mathbb{N}^2$ such that 
\[
d(h) = \left\{ 
\begin{array}{lll}
(0,0) & \text{if} & h=0, \\ 
(0,n) & \text{if} & h=a_{n}x^{n}+\cdots +a_{0},\;\;a_{n}\neq 0, \\ 
(k,n) & \text{if} & h=a_{n}f_{k}^{n}+\cdots +a_{0},\;a_{n}\neq 0,\;
    f_{k}=inv(p_{k}),\\
 & & k\;\;\text{is maximal such that}\;\; f_{k}\precsim f, \\
(k,n) & \text{if} & h=a_{n}f_{k}^{n}+\cdots +a_{0},\;a_{n}\neq 0,\;
    f_{k}=\ing(p_{k}),\;\\
 & & k\;\;\text{is maximal such that}\;\; f_{k}\precsim f, \\
 (k,n) & \text{if} & h=f_{k}^{u}(a_{n}f_{k}^{n}+\cdots+a_{0}),\;
   a_{n}\neq 0,\;a_{0}\neq 0,\;u\in\mathbb{Z},\\
 & & f_{k}=\exp(p_{k}), k\;\;\text{is maximal such that}\;\; f_{k}\precsim f.
\end{array}
\right.
\]
Let $\prec$ be the binary relation over $E\langle f_{1},\ldots ,f_{r}\rangle$ such that
\[ h_1\prec h_2 \Longleftrightarrow k_{1}<k_{2}\vee
(k_{1}=k_{2}\wedge n_{1}<n_{2}) 
\]
where $(k_1,n_1) = d(h_1)$ and $(k_2,n_2) = d(h_2)$. \qed
\end{defn}

It is obvious but crucial to note that $E\langle f_{1},\ldots ,f_{r}\rangle$  
does not contain infinite $\prec$-descending chains. In the following we will 
show that when $ ETSF$ makes a recursive call, the first argument decreases 
with $\prec$. Let $f\in E\langle f_{1},\ldots,f_{r}\rangle$  be such that it 
contains at least one of $f_{1},\ldots,f_{r}$. Let $k$ be maximal such that 
$f_{k}\precsim f$ and let $d(f)=(k,n)$. Note

\begin{enumerate}\itemsep=0.7em
\item[(\ref{r1})]  $ETSF(t;\ f_{1},\ldots,f_{k-1})$.

\noindent 
Note that $t=a_{0}g^{n}+\cdots+a_{n}$ where  $g,a_0,\ldots a_n \in E\langle f_1,\ldots,f_{k-1}\rangle$.\\
Hence  $d(t)=(k',n)$ where $k'< k$. \\
Thus $d(t) \prec d(f)$.

\item[(\ref{r4})]  $ETSF(a_{n};\ f_{1},\ldots,f_{k-1})$.

\noindent 
Note  that $a_n \in E\langle f_1,\ldots,f_{k-1}\rangle$. \\
Hence  $d(a_n)=(k',n')$ where $k'< k$ and $n'$ is a non-negative integer. \\
Thus $d(a_n) \prec d(f)$.

\item[(\ref{r5})]  $ETSF(t;\ f_{1},\ldots,f_{k})$.

\noindent 
Note that $t = c_{i,m+1}f_{k}^{i}+\cdots+c_{0,m+1}$ where $n-1\ge i \ge 0$
and $c_{i,m+1},\ldots, c_{0,m+1} \in E\langle f_1,\ldots,f_{k-1}\rangle$.\\   
Hence if $i>0$ then $d(t)=(k, i)$ else $d(t)=d(c_{0,m+1})=(k',n')$ where 
$k'< k$ and $n'$ is a non-negative integer. \\
Thus $d(t) \prec d(f)$.

\item[(\ref{r2})]  $ETSF(a_{0};\ f_{1},\ldots,f_{k-1})$.

\noindent
Note  that $a_0 \in E\langle f_1,\ldots,f_{k-1}\rangle$. \\
Hence  $d(a_0)=(k',n')$ where $k'< k$ and $n'$ is a non-negative integer. \\
Thus $d(a_0) \prec d(f)$.

\item[(\ref{r3})]  $ETSF(t;\ f_{1},\ldots,f_{k})$. 

\noindent
Note that $t =  c_{n,m+1}f_{k}^{n-i}+\cdots+c_{i,m+1}$ where $n \ge i \ge 1$
and $c_{n,m+1},\ldots, c_{i,m+1} \in E\langle f_1,\ldots,f_{k-1}\rangle$. \\
Hence if $i<n$ then $d(t)=(k, n-i)$ else $d(t)=d(c_{n,m+1})=(k',n')$ where 
$k'< k$ and $n'$ is a non-negative integer. \\
Thus $d(t) \prec d(f)$.
\end{enumerate}

\noindent Hence the algorithm  $ ETSF$  terminates.
Thus  the main algorithm $EISF$ terminates.

\subsection{Partial Correctness}
In this subsection, we prove the partial correctness of the main algorithm 
$EISF$. Let $e \in EI$. 
We need to show that the output of the main algorithm $EISF$ 
is a semi-Fourier sequence of $e$. 
The algorithm makes calls to the subalgorithms~$ET$ and $ETSF$. 
Hence we need to show their partial correctness.
The partial correctness of the subalgorithm $ET$ is obvious.
Thus, we need to show the partial correctness of the subalgorithm~$ETSF$.  

\begin{defn}[Extension]
We say that $g$ \emph{extends} $f$, and write $f\rightsquigarrow g$, if $f$
be transformed into $g$ by applying the commutative ring properties of $+$
and $\cdot$ and by cancellation (replacing $e\cdot \mbox{$\mathrm{inv}$}(e)$
with $1$).
\end{defn}

\begin{example}
$(x^2-1)\mbox{$\mathrm{inv}$}(x-1)\rightsquigarrow x+1$
\end{example}

For the purpose of the proof let us make the following definition.

\begin{defn}[SF sequence] Let $f\in EI$. A sequence
$(g_{1},h_{1}),\ldots,(g_{m},h_{m})$ of pairs of elements of $EI$
is an \emph{SF sequence} for $f$ if 
\begin{enumerate}
\item $f h_{1}\rightsquigarrow g_{1}$
\item $D(g_{k})h_{k+1}\rightsquigarrow g_{k+1}$, for $1\leq k<m$,
\item $D(g_{m})\rightsquigarrow 0$,
\item $h_{k}=t_{k,1}\cdots t_{k,l_{k}}$, for $1\leq k\leq m$, where for each 
$1\leq j\leq l_{k}$ one of the following two conditions holds
   \begin{enumerate}
   \item $t_{i,j}^{-1}\lesssim f$, 
   \item$t_{i,j}=\exp(u)^{-1}$ and $\exp(u)\lesssim e$.
   \end{enumerate}
\end{enumerate}
\end{defn}

It is easy to see that an SF sequence for $f$ is a semi-Fourier sequence 
for $f$.

Let a sequence $e_{1},\ldots,e_{r}$ of all distinct subexpressions
of $e$ of the form $T_{i}(u_{i})$, where $T_{i}\in\{inv,\exp,\ing\}$,
ordered by the non-decreasing value of $rank$. 
Let  
\[
f;f_{1},\ldots,f_{r}= ET(e;e_{1},\ldots,e_{r})
\]
Then we have 
\begin{itemize} 
\item $f_{i}\sim e_{i}$ for $1\leq i\leq r$ 
\item $f\sim e$ 
\item $f\in E\langle f_{1},\ldots,f_{r}\rangle$.
\end{itemize}

We will show that the output of the subalgorithm $ETSF$ is an SF sequence 
for $f$, and therefore a semi-Fourier sequence for $f$.

\begin{enumerate}\itemsep=0.7em

\item[(\ref{o1})] Return $(g_{1},h_{1}),\ldots,(g_{n+1},h_{n+1})$.

\noindent Note that $f\in\mathbb{Q}[x]$. Thus the output is obviously 
an SF sequence of $f$ (in fact the Fourier sequence of $f$).

\item[(\ref{o2})] Return $(g_{1},g^{n}h_{1}),\ldots,(g_{m},h_{m})$.
 
\noindent  Note $f_{k}=\inv(g)$ and $t=a_{n}+\ldots+a_{0}g^{n}$ then 
$d(t)=(k_{1},n_{1})$ with $k_{1}<k$. By inductive hypothesis, 
$(g_{1},h_{1}),\ldots,(g_{m},h_{m})=ETSF(t)$ is an SF sequence for $t$. 
Since 
\begin{equation*}
fg^{n}h_{1}\rightsquigarrow th_{1}\rightsquigarrow g_{1}
\end{equation*}
and $g^{-1}\lesssim f$, 
$(g_{1},g^{n}h_{1}),\ldots,(g_{m},h_{m})$ is an SF sequence for $f$.

\item[(\ref{o5})] If $i=-1$, return $(g_{1},h_{1}),\ldots,(g_{m},h_{m})$.

\noindent 
Note $f_{k}=\ing(g)$ and $d(a_{n})=(k_{1},n_{1})$ with $k_{1}<k$. By
inductive hypothesis, $(e_{1},h_{1}),\ldots,(e_{m},h_{m})=ETSF(a_{n})$ 
is an SF sequence for $a_{n}$. We have%
\begin{equation*}
fh_{1}\sim
a_{n}h_{1}f_{k}^{n}+a_{n-1}h_{1}f_{k}^{n-1}+\ldots+a_{0}h_{1}%
\rightsquigarrow
e_{1}f_{k}^{n}+a_{n-1}h_{1}f_{k}^{n-1}+\ldots+a_{0}h_{1}=g_{1}
\end{equation*}
and, for $1\leq j\leq m$,%
\begin{eqnarray*}
& & D(g_{j}) \\
& \sim &
D(e_{j}f_{k}^{n}+c_{n-1,j}h_{j}f_{k}^{n-1}+\ldots+c_{0,j}h_{j}) \\
& \sim & D(e_{j})f_{k}^{n}+\left(D(c_{n-1,j}h_{j})+e_{j}ng\right)f_{k}^{n-1}+
\\
& &
\left(D(c_{n-2,j}h_{j})+c_{n-1,j}h_{j}(n-1)g\right)f_{k}^{n-2}+\ldots+%
\left(D(c_{0,j}h_{j})+c_{1,j}h_{j}g\right) \\
& \sim &
D(e_{j})f_{k}^{n}+c_{n-1,j+1}f_{k}^{n-1}+c_{n-2,j+1}f_{k}^{n-2}+%
\ldots+c_{0,j+1}
\end{eqnarray*}
Therefore, for $1\leq j<m$, 
\begin{eqnarray*}
D(g_{j})h_{j+1} & \sim &
D(e_{j})h_{j+1}f_{k}^{n}+c_{n-1,j+1}h_{j+1}f_{k}^{n-1}+%
\ldots+c_{0,j+1}h_{j+1} \\
& \rightsquigarrow &
e_{j+1}f_{k}^{n}+c_{n-1,j+1}h_{j+1}f_{k}^{n-1}+\ldots+c_{0,j+1}h_{j+1} \\
& = & g_{j+1}
\end{eqnarray*}
Moreover, since $D(e_{m})\rightsquigarrow0$,%
\begin{equation*}
D(g_{m})\rightsquigarrow c_{n-1,m+1}f_{k}^{n-1}+\ldots+c_{0,m+1}
\end{equation*}
If the answer is returned in step $(\ref{o5})$ then $D(g_{m})\rightsquigarrow0$,
and hence $(g_{1},h_{1}),\ldots,(g_{m},h_{m})$ is an SF sequence
for $f$.

\item[(\ref{o6})] Return $(g_{1},h_{1}),\ldots,(g_{m},h_{m}),(p_{1},q_{1}),\ldots,(p_{l},q_{l})$.

\noindent 
Otherwise
\begin{equation*}
D(g_{m})\rightsquigarrow c_{i,m+1}f_{k}^{i}+\ldots+c_{0,m+1}
\end{equation*}
and, by inductive hypothesis, $(p_{1},q_{1}),\ldots,(p_{l},q_{l})$ 
is an SF sequence for
\[t = c_{i,m+1}f_{k}^{i}+\ldots+c_{0,m+1}\]
Hence
\begin{equation*}
D(g_{m})q_{1}\rightsquigarrow t q_{1}\rightsquigarrow p_{1}
\end{equation*}
and therefore $(g_{1},h_{1}),\ldots,(g_{m},h_{m}),(p_{1},q_{1}),%
\ldots,(p_{l},q_{l})$ is an SF sequence for~$f$.

\item[(\ref{o3})] If $i=n+1$, return $(g_{1},f_{k}^{-u}h_{1}),\ldots,(g_{m},h_{m})$.

\noindent Note $f_{k}=\exp(g)$ and $d(a_{0})=(k_{1},n_{1})$ with $k_{1}<k$. 
By inductive hypothesis, 
$(e_{1},h_{1}),\ldots,(e_{m},h_{m})=ETSF(a_{0})$ 
is an SF sequence for $a_{0}$. We have%
\begin{equation*}
f f_{k}^{-u} h_{1} \rightsquigarrow
a_{n}h_{1}f_{k}^{n}+\ldots+a_{1}h_{1}f_{k}+a_{0}h_{1}\rightsquigarrow
a_{n}h_{1}f_{k}^{n}+\ldots+a_{1}h_{1}f_{k}+e_{1}=g_{1}
\end{equation*}
and, for $1\leq j\leq m$,%
\begin{eqnarray*}
&      & D(g_{j})  \\
& \sim &
D(c_{n,j}h_{j}f_{k}^{n})+\ldots+D(c_{1,j}h_{j}f_{k})+D(e_{j}) \\
& \sim &
\left(D(c_{n,j}h_{j})+c_{n,j}h_{j}nD(g)\right)f_{k}^{n}+\ldots+%
\left(D(c_{1,j}h_{j})+c_{1,j}h_{j}D(g)\right)+D(e_{j}) \\
& \sim & c_{n,j+1}f_{k}^{n}+\ldots+c_{1,j+1}f_{k}+D(e_{j})
\end{eqnarray*}
Therefore, for $1\leq j<m$, 
\begin{eqnarray*}
D(g_{j})h_{j+1} & \sim &
c_{n,j+1}h_{j+1}f_{k}^{n}+\ldots+c_{1,j+1}h_{j}f_{k}+D(e_{j})h_{j} \\
& \rightsquigarrow &
c_{n,j+1}h_{j+1}f_{k}^{n}+\ldots+c_{1,j+1}h_{j}f_{k}+e_{j+1} \\
& = & g_{j+1}
\end{eqnarray*}
Moreover, since $D(e_{m})\rightsquigarrow0$,%
\begin{equation*}
D(g_{m})\rightsquigarrow c_{n,m+1}f_{k}^{n}+\ldots+c_{1,m+1}f_{k}
\end{equation*}
If $i=n+1$ then $D(g_{m})\rightsquigarrow0$. Since $f_{k}=\exp(g)\lesssim f$,
$(g_{1},f_{k}^{-u}h_{1}),\ldots,(g_{m},h_{m})$ is an SF sequence for~$f$.

\item[(\ref{o4})] Return $(g_{1},f_{k}^{-u}h_{1}),\ldots,(g_{m},h_{m}),(p_{1},f_{k}^{-i}q_{1}),\ldots,(p_{l},q_{l})$.
 
\noindent Otherwise 
\begin{equation*}
D(g_{m})\rightsquigarrow\left(c_{n,m+1}f_{k}^{n-i}+\ldots+c_{i,m+1}%
\right)f_{k}^{i}
\end{equation*}
and, by inductive hypothesis, $(p_{1},q_{1}),\ldots,(p_{l},q_{l})$ 
is an SF sequence for
\[t = c_{n,m+1}f_{k}^{n-i}+\ldots+c_{i,m+1}\]
Hence
\begin{eqnarray*}
D(g_{m})f_{k}^{-i}q_{1} & \rightsquigarrow &
t f_{k}^{i}f_{k}^{-i}q_{1} \\
& \rightsquigarrow &  t q_{1}
\\
& \rightsquigarrow & p_{1}
\end{eqnarray*}
and therefore $(g_{1},f_{k}^{-u}h_{1}),\ldots,(g_{m},h_{m}),(p_{1},f_{k}^{-i}q_{1}),%
\ldots,(p_{l},q_{l})$ is an SF sequence for $f$.
\end{enumerate}

\section{Examples}
\label{examples}
\begin{example}[exp \cite{S1}] 
Find a semi-Fourier sequence of the function
\[
f = e^{e^{e^x}} e^{-e^{e^{x-e^{-e^x}}}}-10^5\] \par 
\noindent \sf{Input expression:} \hfill \par 
$e=\exp (\exp (\exp (x))) \exp (-\exp (\exp (x-\exp (-\exp (x)))))-100000$ \par

\break

\noindent \sf{Output:} \hfill \par 

$
\begin{array}[t]{r|l|l} 
i & g_i & h_i \\ 
\hline 
1 & f_5 f_7-100000 & 1 \\ 
2 & \left(-f_2-f_1^{-1}\right) f_3^{-1} f_4 f_6+1 & f_1^{-1} f_3^{-1} f_5^{-1} f_7^{-1} \\ 
3 & \left(-\left(f_1 f_2^2\right)+\cdots -f_1^{-1}\right) f_4-f_1 f_2^2+\cdots +1 & f_3 f_4^{-1} f_6^{-1} \\ 
4 & \left(-\left(f_1 f_2^2\right)+\cdots -3 f_1^{-1}+2\right) f_4+\left(2 f_1-1\right) f_2-2 f_1+4 & f_1^{-1} f_2^{-1} \\ 
5 & \left(-\left(f_1 f_2^3\right)+\cdots +2 f_1^{-1}\right) f_4+\left(-2 f_1+3\right) f_2-2 & f_1^{-1} \\ 
6 & \left(-\left(f_1 f_2^3\right)+\cdots +2 f_1+19 f_1^{-1}-14\right) f_4+2 f_1-5 & f_1^{-1} f_2^{-1} \\ 
7 & \left(-\left(f_1 f_2^4\right)+\cdots -14 f_1^{-1}+4\right) f_4+2 & f_1^{-1} \\ 
8 & -\left(f_1^2 f_2^5\right)+\cdots +4 & f_4^{-1} \\ 
9 & \left(5 f_1^2-2 f_1\right) f_2^4+\cdots +12 f_1^2+\cdots +232 & f_1^{-1} f_2^{-1} \\ 
10 & \left(-20 f_1^2+\cdots -2\right) f_2^4+\cdots +24 f_1-116 & f_1^{-1} \\ 
11 & \left(80 f_1^2+\cdots +26\right) f_2^4+\cdots +24 & f_1^{-1} \\ 
12 & \left(-320 f_1^2+\cdots -216\right) f_2^3+\cdots +120 f_1^2+\cdots +3032 & f_1^{-1} f_2^{-1} \\ 
13 & \left(960 f_1^2+\cdots +1256\right) f_2^3+\cdots +240 f_1-1304 & f_1^{-1} \\ 
14 & \left(-2880 f_1^2+\cdots -6232\right) f_2^3+\cdots +240 & f_1^{-1} \\ 
15 & \left(8640 f_1^2+\cdots +28008\right) f_2^2+\cdots +1296 f_1^2+\cdots +38776 & f_1^{-1} f_2^{-1} \\ 
16 & \left(-17280 f_1^2+\cdots -89712\right) f_2^2+\cdots +2592 f_1-15168 & f_1^{-1} \\ 
17 & \left(34560 f_1^2+\cdots +264096\right) f_2^2+\cdots +2592 & f_1^{-1} \\ 
18 & \left(-69120 f_1^2+\cdots -732096\right) f_2+12096 f_1^2+\cdots +409392 & f_1^{-1} f_2^{-1} \\ 
19 & \left(69120 f_1^2+\cdots +1209024\right) f_2+24192 f_1-149472 & f_1^{-1} \\ 
20 & \left(-69120 f_1^2+\cdots -1824192\right) f_2+24192 & f_1^{-1} \\ 
21 & 69120 f_1^2+\cdots +2577600 & f_1^{-1} f_2^{-1} \\ 
22 & 138240 f_1-891648 & f_1^{-1} \\ 
23 & 138240 & f_1^{-1} \\ 
\end{array}
$

\noindent where

$
\begin{array}[t]{lll}
f_1 &=& e^x \\
f_2 &=& e^{-f_1} \\
f_3 &=& e^{f_1} \\
f_4 &=& e^{x-f_2} \\
f_5 &=& e^{f_3} \\
f_6 &=& e^{f_4} \\
f_7 &=& e^{-f_6}
\end{array}
$

\end{example}

\begin{example}[exp-inv \cite{S1}]
Find a semi-Fourier sequence of the function
\[
f = e^x \left(e^{\frac{1}{x}-e^{-x}}-e^{\frac{1}{x}}\right)+5
\] \par 
\noindent \sf{Input expression:} \hfill \par 
$e=\exp (x) (\exp (\inv(x)-\exp (-x))-\exp (\inv(x)))+5$ \par
\noindent \sf{Output:} \hfill \par 

$
\begin{array}[t]{r|l|l} 
i & g_i & h_i \\ 
\hline 
1 & f_2 f_5-f_2 f_4+5 & 1 \\ 
2 & \left(-\left(x^2 f_3^2\right)+x^2 f_1+x^2\right) f_4^{-1} f_5-x^2+1 & f_2^{-1} f_3^{-2} f_4^{-1} \\ 
3 & \left(2 \left(x^2 f_3^3\right)+\cdots +x^2 f_1^2+\cdots +2 x\right) f_4^{-1} f_5-2 x & 1 \\ 
4 & \left(-6 \left(x^2 f_3^4\right)+\cdots +x^2 f_1^3+\cdots +2\right) f_4^{-1} f_5-2 & 1 \\ 
5 & x^7 f_1^3+\cdots -x^5 & f_1^{-1} f_3^{-5} f_4 f_5^{-1} \\ 
6 & \left(-3 x^7+7 x^6\right) f_1^3+\cdots -5 x^4 & 1 \\ 
7 & \left(9 x^7+\cdots +42 x^5\right) f_1^3+\cdots -20 x^3 & 1 \\ 
8 & \left(-27 x^7+\cdots +210 x^4\right) f_1^3+\cdots -60 x^2 & 1 \\ 
9 & \left(81 x^7+\cdots +840 x^3\right) f_1^3+\cdots -120 x & 1 \\ 
10 & \left(-243 x^7+\cdots +2520 x^2\right) f_1^3+\cdots -120 & 1 \\ 
11 & \left(729 x^7+\cdots +5040 x\right) f_1^2+\cdots +4 x^7+\cdots -15120 & f_1^{-1} \\ 
12 & \left(-1458 x^7+\cdots +5040\right) f_1^2+\cdots +28 x^6+\cdots +88200 & 1 \\ 
13 & \left(2916 x^7+\cdots -110880\right) f_1^2+\cdots +168 x^5+\cdots -272160 & 1 \\ 
14 & \left(-5832 x^7+\cdots +1285200\right) f_1^2+\cdots +840 x^4+\cdots +491400 & 1 \\ 
15 & \left(11664 x^7+\cdots -10503360\right) f_1^2+\cdots +3360 x^3+\cdots -539280 & 1 \\ 
16 & \left(-23328 x^7+\cdots +68216400\right) f_1^2+\cdots +10080 x^2+\cdots +355320 & 1 \\ 
17 & \left(46656 x^7+\cdots -375732000\right) f_1^2+\cdots +20160 x-129600 & 1 \\ 
18 & \left(-93312 x^7+\cdots +1827226800\right) f_1^2+\cdots +20160 & 1 \\ 
19 & \left(186624 x^7+\cdots -8060371200\right) f_1-320 x^7+\cdots +1040048640 & f_1^{-1} \\ 
20 & \left(-186624 x^7+\cdots +24812928000\right) f_1-2240 x^6+\cdots -1165919760 & 1 \\ 
21 & \left(186624 x^7+\cdots -68012179200\right) f_1-13440 x^5+\cdots +994933440 & 1 \\ 
22 & \left(-186624 x^7+\cdots +168875965440\right) f_1-67200 x^4+\cdots -635079360 & 1 \\ 
23 & \left(186624 x^7+\cdots -385341707520\right) f_1-268800 x^3+\cdots +293706240 & 1 \\ 
24 & \left(-186624 x^7+\cdots +817742822400\right) f_1-806400 x^2+\cdots -92962560 & 1 \\ 
25 & \left(186624 x^7+\cdots -1630128648960\right) f_1-1612800 x+18017280 & 1 \\ 
26 & \left(-186624 x^7+\cdots +3078168468480\right) f_1-1612800 & 1 \\ 
27 & 186624 x^7+\cdots -5544580204800 & f_1^{-1} \\ 
28 & 1306368 x^6+\cdots +4038444184320 & 1 \\ 
29 & 7838208 x^5+\cdots -2349935400960 & 1 \\ 
30 & 39191040 x^4+\cdots +1065791623680 & 1 \\ 
31 & 156764160 x^3+\cdots -362821939200 & 1 \\ 
32 & 470292480 x^2+\cdots +87160872960 & 1 \\ 
33 & 940584960 x-13168189440 & 1 \\ 
34 & 940584960 & 1 \\ 
\end{array}
$

\noindent where

$
\begin{array}[t]{lll}
f_1 &=& e^{-x} \\
f_2 &=& e^x \\
f_3 &=& \frac{1}{x} \\
f_4 &=& e^{f_3} \\
f_5 &=& e^{f_3-f_1}
\end{array}
$

\end{example}

\begin{example}[exp-log (toy)]
Find a semi-Fourier sequence of the function
\[
f =  e^x \log x + e^{x^2}+x
\] \par 
\noindent \sf{Input expression:} \hfill \par 
$ \exp(x)\  \ing(\inv(x)) +\exp(x^2)+x$ \par
\noindent \sf{Output:} \hfill \par 
$
\begin{array}[t]{r|l|l} 
i & g_i & h_i \\ 
\hline 
1 & f_4+f_1^{-1} f_2+x f_1^{-1} & f_1^{-1} \\ 
2 & \left(2 x^2-x\right) f_2+f_1-x^2+x & x f_1 \\ 
3 & \left(4 x^3+\cdots -1\right) f_2+f_1-2 x+1 & 1 \\ 
4 & \left(8 x^4+\cdots +4\right) f_2+f_1-2 & 1 \\ 
5 & \left(16 x^5+\cdots -6\right) f_1^{-1} f_2+1 & f_1^{-1} \\ 
6 & 32 x^6+\cdots +54 & f_1 f_2^{-1} \\ 
7 & 192 x^5+\cdots -108 & 1 \\ 
8 & 960 x^4+\cdots +672 & 1 \\ 
9 & 3840 x^3+\cdots -912 & 1 \\ 
10 & 11520 x^2+\cdots +5568 & 1 \\ 
11 & 23040 x-3840 & 1 \\ 
12 & 23040 & 1 \\ 
\end{array}
$

\noindent where

$
\begin{array}[t]{lll}
f_1 &=& e^x \\
f_2 &=& e^{x^2} \\
f_3 &=& \frac{1}{x} \\
f_4 &=& \log x
\end{array}
$\end{example}

\begin{example}[exp-log \cite{S1}]
Find a semi-Fourier sequence of the function
\[
f = \frac{x}{\exp (x)-1}-\log (1-\exp (-x))-\frac{1}{x}
\] \par 
\noindent \sf{Input expression:} \hfill \par 
$x \inv(\exp (x)-1)-\ing(\inv(\exp (x)-1))-\inv(x)$ \par

\break

\noindent \sf{Output:} \hfill \par 

$
\begin{array}[t]{r|l|l} 
i & g_i & h_i \\ 
\hline 
1 & -f_4+x f_3-f_2 & 1 \\ 
2 & f_1^2+\left(-x^3-2\right) f_1+1 & f_2^{-2} f_3^{-2} \\ 
3 & 2 f_1-x^3-3 x^2-2 & f_1^{-1} \\ 
4 & 2 f_1-3 x^2-6 x & 1 \\ 
5 & 2 f_1-6 x-6 & 1 \\ 
6 & 2 f_1-6 & 1 \\ 
7 & 2 & f_1^{-1} \\ 
\end{array}
$

\noindent where

$
\begin{array}[t]{lll}
f_1 &=& e^x \\
f_2 &=& \frac{1}{x} \\
f_3 &=& -\frac{1}{1-f_1} \\
f_4 &=& \log \left(1-e^{-x}\right)
\end{array}
$

\end{example}

\begin{example}[exp-log-arccos \cite{S2}]
Find a semi-Fourier sequence of the function
\[
f = 2 \arccos\left(\frac{x}{2}\right)-\frac{\pi }{2}-\frac{1}{2} x \sqrt{4-x^2}
\] \par 
\noindent \sf{Input expression:} \hfill \par 
$2\, \ing\left(-\exp \left(-\frac{1}{2}\, \ing\left(-2\, x\, \inv\left(4-x^2\right)\right)\right)\right)-\frac{1}{2} \, x \exp \left(\frac{1}{2}\, \ing\left(-2 \, x \, \inv\left(4-x^2\right)\right)\right)$ \par
\noindent \sf{Output:} \hfill \par 

$
\begin{array}[t]{r|l|l} 
i & g_i & h_i \\ 
\hline 
1 & 2 f_5-\frac{1}{2} \left(x f_4\right) & 1 \\ 
2 & \left(\frac{1}{2} \left(x^2 f_1\right)-\frac{1}{2}\right) f_3^{-1} f_4-2 & f_3^{-1} \\ 
3 & 2 x & f_1^{-1} f_3 f_4^{-1} \\ 
4 & 2 & 1 \\ 
\end{array}
$

\noindent where

$
\begin{array}[t]{lll}
f_1 &=& \frac{1}{4-x^2} \\
f_2 &=& \log \left(4-x^2\right) \\
f_3 &=& e^{-\frac{f_2}{2}} \\
f_4 &=& e^{\frac{f_2}{2}} \\
f_5 &=& \arccos\left(\frac{x}{2}\right)-\frac{\pi }{4}
\end{array}
$

\end{example}

\begin{example}[exp-int]
Find a semi-Fourier sequence of the function
\[
f =\Ei\left(x^2+2\right)-\, \li\left(x^2+2\right)-e^{x^2+2} 
\] \par 
\noindent \sf{Input expression:} \hfill \par 
$\ing\left(2 x \exp \left(x^2+2\right) \, \inv\left(x^2+2\right)\right)-\exp \left(x^2+2\right)-\, \ing\left(2 x \, \inv\left(\, \
\ing\left(2 x \, \inv\left(x^2+2\right)\right)\right)\right)$\par

\break

\noindent \sf{Output:} \hfill \par 

$
\begin{array}[t]{r|l|l} 
i & g_i & h_i \\ 
\hline 
1 & -f_ 6+f_ 4-f_ 1 & 1 \\ 
2 & \left(-2 x^3-2 x \right) f_ 3+\left(-2 x^3-4 x \right) f_ 1^{-1} & \left(x^2+2\right) f_ 1^{-1} f_ 3 \\ 
3 & \left(-6 x^2-2\right) f_ 3+\left(-4 x^4-4 x^2\right) f_ 2+\left(4 x^4+\cdots -4\right) f_ 1^{-1} & 1 \\ 
4 & -12 \left(x f_ 3\right)+\left(8 x^5+8 x^3\right) f_ 2^2+\cdots +\left(-8 x^5+\cdots +12 x \right) f_ 1^{-1} & 1 \\ 
5 & -12 f_ 3+\left(-32 x^6-32 x^4\right) f_ 2^3+\cdots +\left(16 x^6+\cdots +12\right) f_ 1^{-1} & 1 \\ 
6 & \left(-24 x^7+\cdots -1440 x \right) f_ 1-32 x^{15}+\cdots -4480 x^3 & \left(x^8+\cdots +16\right) f_ 1 \\ 
7 & \left(-48 x^8+\cdots -1440\right) f_ 1-480 x^{14}+\cdots -13440 x^2 & 1 \\ 
8 & \left(-96 x^9+\cdots -12480 x \right) f_ 1-6720 x^{13}+\cdots -26880 x & 1 \\ 
9 & \left(-192 x^{10}+\cdots -12480\right) f_ 1-87360 x^{12}+\cdots -26880 & 1 \\ 
10 & \left(-384 x^{11}+\cdots -187200 x \right) f_ 1-1048320 x^{11}+\cdots -645120 x & 1 \\ 
11 & \left(-768 x^{12}+\cdots -187200\right) f_ 1-11531520 x^{10}+\cdots -645120 & 1 \\ 
12 & \left(-1536 x^{13}+\cdots -3951360 x \right) f_ 1-115315200 x^9+\cdots -322560 x & 1 \\ 
13 & \left(-3072 x^{14}+\cdots -3951360\right) f_ 1-1037836800 x^8+\cdots -322560 & 1 \\ 
14 & \left(-6144 x^{15}+\cdots -105477120 x \right) f_ 1-8302694400 x^7+\cdots +766402560 x & 1 \\ 
15 & \left(-12288 x^{16}+\cdots -105477120\right) f_ 1-58118860800 x^6+\cdots +766402560 & 1 \\ 
16 & \left(-24576 x^{17}+\cdots -3353011200 x \right) f_ 1-348713164800 x^5+\cdots +29698099200 x & 1 \\ 
17 & \left(-49152 x^{18}+\cdots -3353011200\right) f_ 1-1743565824000 x^4+\cdots +29698099200 & 1 \\ 
18 & \left(-98304 x^{19}+\cdots -122464742400 x \right) f_ 1-6974263296000 x^3-199264665600 x & 1 \\ 
19 & \left(-196608 x^{20}+\cdots -122464742400\right) f_ 1-20922789888000 x^2-199264665600 & 1 \\ 
20 & \left(-393216 x^{21}+\cdots -5023130112000 x \right) f_ 1-41845579776000 x & 1 \\ 
21 & \left(-786432 x^{22}+\cdots -5023130112000\right) f_ 1-41845579776000 & 1 \\ 
22 & -1572864 x^{23}+\cdots -227809328947200 x & f_ 1^{-1} \\ 
23 & -36175872 x^{22}+\cdots -227809328947200 & 1 \\ 
\vdots & \vdots  & \vdots \\ 
45 & -40661706455989579897896960000 & 1 \\ 
\end{array}
$

\noindent where

$
\begin{array}[t]{lll}
f &=& -f_ 6+f_ 4-f_ 1\\
f_ 1 &=& e^{x^2+2} \\
f_ 2 &=& \frac{1}{x^2+2} \\
f_ 3 &=& \log \left(x^2+2\right) \\
f_ 4 &=& \, \Ei\left(x^2+2\right) \\
f_ 5 &=& \frac{1}{f_ 3} \\
f_ 6 &=& \, \li\left(x^2+2\right)
\end{array}$

\end{example}

\bibliographystyle{plain}
\bibliography{SF_ExpInt}

\end{document}